# Nonlinear spectra of spinons and holons in short GaAs quantum wires

M. Moreno[1], C.J.B. Ford[1], Y. Jin[1], J.P. Griffiths[1], I. Farrer[1,†], G.A.C. Jones[1], D.A. Ritchie[1], O. Tsyplyatyev[2,†] & A.J. Schofield[2]

One-dimensional electronic fluids are peculiar conducting systems, where the fundamental role of interactions leads to exotic, emergent phenomena, such as spin-charge (spinon-holon) separation. The distinct low-energy properties of these 1D metals are successfully described within the theory of linear Luttinger liquids, but the challenging task of describing their high-energy nonlinear properties has long remained elusive. Recently, novel theoretical approaches accounting for nonlinearity have been developed, yet the rich phenomenology that they predict remains barely explored experimentally. Here, we probe the nonlinear spectral characteristics of short GaAs quantum wires by tunnelling spectroscopy, using an advanced device consisting of 6000 wires. We find evidence for the existence of an inverted (spinon) shadow band in the main region of the particle sector, one of the central predictions of the new nonlinear theories. A (holon) band with reduced effective mass is clearly visible in the particle sector at high energies.

[1] Cavendish Laboratory, University of Cambridge, JJ Thomson Avenue, Cambridge CB3 0HE, UK. [2] School of Physics and Astronomy, University of Birmingham, Edgbaston, Birmingham B15 2TT, UK. † Present addresses: Department of Electronic & Electrical Engineering, University of Sheffield, Mappin Street, Sheffield S1 3JD, UK (I.F.); Institut für Theoretische Physik, Universität Frankfurt, Max-von-Laue Strasse 1, 60438 Frankfurt, Germany (O.T.). Correspondence and requests for materials should be addressed to M.M. (email: mmoreno.spain@gmail.com).

The electronic properties of one-dimensional (1D) electron systems are fundamentally different from those of their two-dimensional (2D) and three-dimensional (3D) counterparts, owing to the prominent role of interactions under 1D confinement[1–9]. Fermi-liquid theory, applicable to normal 2D and 3D electron systems, breaks down spectacularly in the 1D case, which is better described by the theory of Tomonaga–Luttinger liquids[10,11]. While elementary excitations of Fermi liquids behave as quasi-free fermions, those of interacting electrons in 1D systems are collective bosonic modes. Out of equilibrium, electrons confined to 1D lose their individual identity, separating into two collective excitation types (quantized waves of density): spin modes (spinons) that carry spin without charge, and charge modes (holons) that carry charge without spin[2,12]. Spinons and holons travel at different velocities, resulting in so-called spin-charge separation, revealed in photoemission and tunnelling experiments[12–15]. The spectral function $A(\hbar \mathbf{k}, \hbar\omega)$, describing the probability for an electron with momentum $\hbar \mathbf{k}$ and energy $\hbar\omega > 0$ ($\hbar\omega < 0$) to tunnel into (out of) the system, theoretically displays narrow Lorentzian singularities for normal 3D and 2D systems, whereas it displays wide asymmetric power-law singularities for 1D systems[12,16,17].

The theory of Tomonaga–Luttinger liquids, based on a linearization of the dispersion relation around the right ($+k_F^{1D}$) and left ($-k_F^{1D}$) Fermi points[10,11], has been successfully used for a long time to describe 1D electron systems in the limit of low energies but, in order to understand their behaviour at higher energies, that is, away from the Fermi points, the curvature of the dispersion relation has to be taken into account. Recently, a new nonlinear theory of 1D quantum fluids beyond the low-energy limit has been developed[18–28]. The new theory represents a giant advance towards full understanding of the behaviour of 1D electron systems, but many of its specific predictions remain barely explored experimentally.

Here, we report on the spectra of elementary excitations in short GaAs quantum wires, probed by momentum- and energy-resolved tunnelling spectroscopy, focusing on the nonlinear high-energy regime. We use a 1D–2D vertical tunnelling device with a large number of wires in parallel, to boost the wire signal and so maximize the chances to observe weak features of the wire spectral function. Our tunnelling conductance maps reveal, for the first time, the existence of an inverted (spinon) shadow band in the main region of the particle sector, symmetrically replicating the dispersion of the main spinon hole band, as anticipated by the nonlinear theory. They also reveal a (holon) band with reduced effective mass in the particle sector at high energies. Holons appear to be long-lived in the particle sector, but short-lived in the hole sector.

## Results

**Vertical tunnelling device.** Our device (Fig. 1) consists of a double GaAs quantum-well structure separated by a thin AlGaAs barrier, and various surface gates, among them an array of 1 µm-long wire gates (WG) interconnected with air bridges (Fig. 2). We apply a negative voltage $V_{wg}$ to the wire gates, which is strong enough to pinch off (fully deplete) the upper-well (UW) regions under the wire gates, but not strong enough to pinch off the lower-well (LW) regions under the wire gates. Hence, in the lower well, electrons can move under the wire gates, but in the upper well they cannot. Electrons in the upper well are laterally confined to a network of narrow parallel channels (W regions in Fig. 1c) so-called wires, and perpendicular wide trenches (La, Lb, Lc and Ld regions in Fig. 1c, jointly denoted L) so-called leads. In the UW-wire regions, electrons are strongly confined to a narrow width, thus becoming 1D. In contrast, in the UW-lead regions, no dimension is very narrow, and so electrons remain 2D. In the lower well, electrons are not laterally confined, and so their dimensionality is 2D. The quantum wells are separately contacted using a surface-gate depletion technique (see Methods section for details). A fixed ac voltage $V_{ac} = 50\,\mu V$ and a variable dc-bias $V_{dc}$ are applied between them. The bias is applied to the lower well, while the upper well is grounded. The current $I$, injected through one of the quantum wells, is forced to tunnel through the barrier that separates them, before leaving the device through the other quantum well. The differential tunnelling conductance $G = dI/dV_{ac}$, in a variable in-plane magnetic field $B$ perpendicular to the wires, is measured at ∼ 57 mK lattice temperature (higher electron temperature) in a dilution refrigerator, using standard lock-in techniques. The W and L regions of the device contribute distinct signals in $G(B, V_{dc})$ maps. The W signal has 1D–2D nature, that is, it corresponds to tunnelling between the UW 1D wires and the LW 2D electron gas. The L signal has 2D–2D nature, that is, it corresponds to tunnelling between the UW 2D leads and the LW 2D electron gas.

**Control of electron density in the wires.** The electron density in the UW wires can be controlled by tuning the wire–gate voltage $V_{wg}$. This is shown in Figs 3 and 4. Figure 3 maps $dG/dV_{wg}$ for conditions close to equilibrium ($V_{dc} \approx 0$). The map reveals the 1D wire subbands participating in 1D–2D tunnelling. The single-subband regime is achieved with $V_{wg}$ in the range $-0.78\,V \lesssim V_{wg} \lesssim -0.69\,V$. For $V_{wg} > -0.69\,V$, more than one wire subband participates in 1D–2D tunnelling and, for $V_{wg} < -0.78\,V$, the wires are completely depleted of carriers. Figure 3 shows the magnetic-field values $B_W^-$, $B_W^+$, $-B_L^-$ and $B_L^+$, at which tunnelling resonances cross the $V_{dc} = 0$ axis in $G(B, V_{dc})$ maps (for example Fig. 5), as a function of the wire–gate voltage. There are two sets of crossing values. One, $\pm B_W^\pm$, corresponds to tunnelling in the W regions of the device, and another, $\pm B_L^\pm$, to tunnelling in the L regions. Note that $B_W^\pm$ denote the values corresponding to the first UW wire subband. The crossing values are related to the Fermi-wavevector components along the wire direction as follows:

$$B_W^\pm = \hbar\left(k_F^{LW,W} \pm k_F^{1D}\right)/(ed) \quad (1)$$

$$B_L^\pm = \hbar\left(k_F^{LW,L} \pm k_F^{UW,L}\right)/(ed). \quad (2)$$

Here, $\hbar k_F^{1D}$ ($\hbar k_F^{LW,W}$) is the Fermi-wavevector component in the W regions of the upper well (lower well), $\hbar k_F^{UW,L}$ ($\hbar k_F^{LW,L}$) is the Fermi-wavevector component in the L regions of the upper well (lower well), $d$ is the distance between the centres of the UW and LW wavefunctions, and $-e$ is the electron charge. It happens that $k_F^{1D} < k_F^{LW,W} \approx k_F^{LW,L} < k_F^{UW,L}$, and so $B_W^- > 0$ and $B_L^- < 0$. The Fermi wavenumber $k_F^{1D}$ ($k_F^{2D}$) in a 1D (2D) system is related to its equilibrium 1D (2D) electron density $n_{1D}$ ($n_{2D}$) as follows:

$$n_{1D} = 2k_F^{1D}/\pi \quad (3)$$

$$n_{2D} = (k_F^{2D})^2/(2\pi). \quad (4)$$

Hence, the equilibrium electron densities in the upper and lower wells, for the wire (W) and lead (L) regions of the device, can be estimated from the values of $B_W^\pm$ and $B_L^\pm$ in Fig. 3. These are shown in Fig. 4. The wire gates efficiently tune electron density in the wire (W) regions of the upper well (Fig. 4a), but have little influence on densities in the lower well and in the lead (L) regions of the upper well (Fig. 4b). Note that the LW density in the W regions is only slightly smaller than that in the L regions (Fig. 4b), and notably higher than the UW density in the W regions (Fig. 4a).

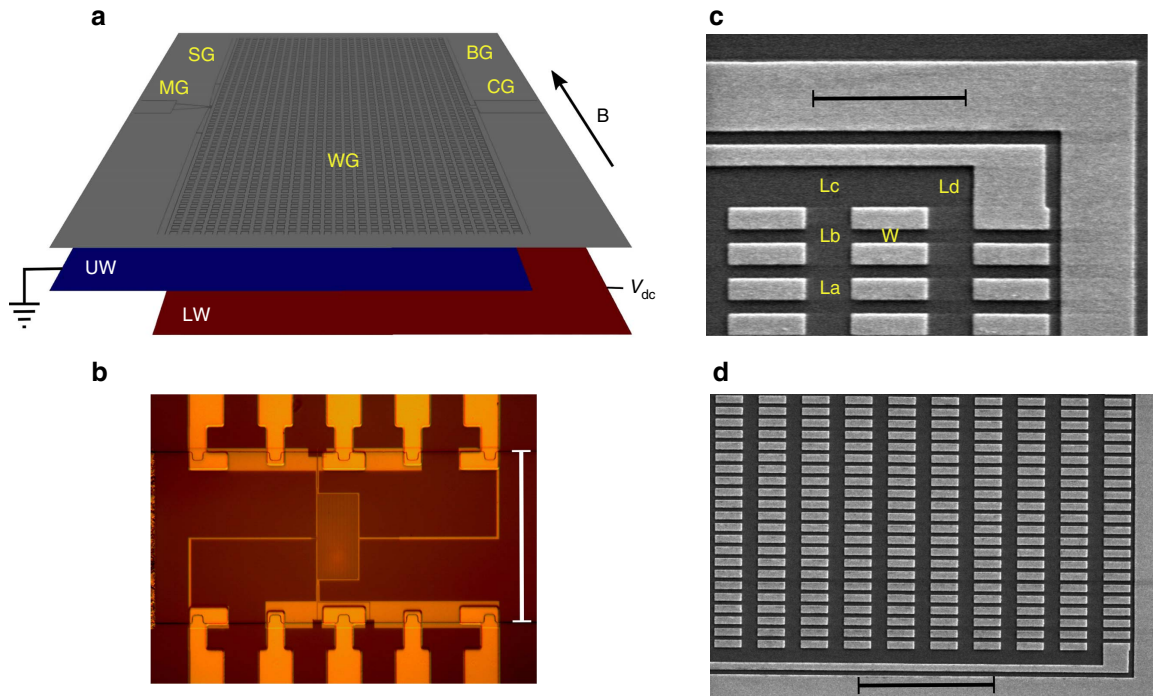

**Figure 1 | Tunnelling device.** (**a**) Schematic of the tunnelling device. It consists of a double quantum-well structure and various surface gates: split (SG), mid-line (MG), bar (BG) and cut-off (CG) gates—used to define the experimental area and to set-up tunnelling conditions—and a 30 × 200 array of parallel 1 μm-long air-bridged wire gates—used to define the quantum wires in the upper well. A dc-bias $V_{dc}$ is applied to the lower well (LW), while the upper well (UW) is grounded. An in-plane magnetic field $B$ is applied perpendicular to the long side of the wire gates. (**b**) Optical micrograph of the surface gates, fabricated on a Hall bar 200 μm-wide (white scale bar). (**c**,**d**) Scanning electron microscopy (SEM) micrographs of the upper-right (**c**) and lower-right (**d**) corners of the experimental area (device surface) before bridge fabrication. In **c** the black scale bar is 2 μm-long, and in **d** it is 5 μm-long. Dark grey corresponds to (oxidized) semiconductor areas, and bright grey to metallic gates. Electron fluids lie in the (buried) upper and lower wells. In the lower well, electrons are not laterally confined; they can move under the wire gates. In the upper well, electrons cannot move under the wire gates; they are confined to a network of narrow wires (W regions in **c**) and wide leads (La, Lb, Lc and Ld regions in **c**, jointly denoted L). UW electrons have 1D character in W regions, and 2D character in L regions. LW electrons have 2D character.

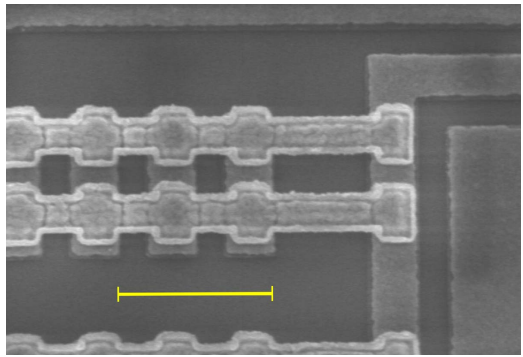

**Figure 2 | Air bridges.** Scanning electron microscopy (SEM) micrograph of a device's surface region, showing air-bridge interconnections between wire gates. Dark grey corresponds to (oxidized) semiconductor areas, and bright grey to metallic gates. The yellow scale bar is 1 μm-long.

**Regime of a single 1D wire subband filled.** The rate of electron tunnelling between the upper and lower wells depends on the overlap between the respective spectral functions, which can be varied by tuning the magnetic field and the dc-bias. The magnetic field $B$ and the dc-bias $V_{dc}$ shift the UW and LW spectral functions relative to each other, in the momentum and energy directions, by $\hbar\Delta k = eBd$ and $\hbar\Delta\omega = eV_{dc}$, respectively. Hence, by mapping out the tunnelling conductance $G(B, V_{dc})$, the spectral characteristics of the UW wires, UW leads and lower well are probed. Figure 5 shows a map corresponding to the single-subband regime ($V_{wg} = -0.69$ V). It displays quasi-parabolic resonances, corresponding to tunnelling from ground states of the source (spectrometer) to excited states of the drain (probe) occurring with conservation of energy and momentum, that is, the tunnelling condition

$$\varepsilon_{UW}(k - \Delta k) = \varepsilon_{LW}(k) - eV_{dc} \quad (5)$$

is satisfied, with either $|k| = k_F^{LW}$ or $|k - \Delta k| = k_F^{UW}$, where $k_F^{LW}$ ($k_F^{UW}$) is the Fermi wavenumber of the lower well (upper well) in the wire or lead regions of the device. Here, $\varepsilon_{UW}(k)$ and $\varepsilon_{LW}(k)$ are the dispersions of the elementary excitations in the upper and lower wells, respectively, in the wire or lead regions. The map reflects tunnelling events occurring between ground states of the upper well and excited states of the lower well, and between ground states of the lower well and excited states of the upper well. The wire (W) and lead (L) regions of the device contribute distinct signals. The solid (dashed) green lines in Fig. 5 mark resonances corresponding to tunnelling in W (L) regions, from UW ground states $\varepsilon_{UW}(\pm k_F^{UW})$ to LW states; they reveal the LW dispersion in W (L) regions of the device. The dash-dotted green lines mark resonances corresponding to tunnelling in L regions from LW ground states $\varepsilon_{LW}(\pm k_F^{LW,L})$ to UW states; they reveal the UW-lead dispersion. Most importantly here, the resonances marked by the black and white arrows correspond to tunnelling in W regions, from LW ground states $\varepsilon_{LW}(\pm k_F^{LW,W})$ to UW-wire states. They reveal 1D UW-wire dispersions: a holon band in the particle sector ($h+$), and a spinon band in the hole sector ($s-$), as elucidated below. Remarkably, an extra wire band ($s+$), symmetric of $s-$, is also observed in the wires' particle sector.

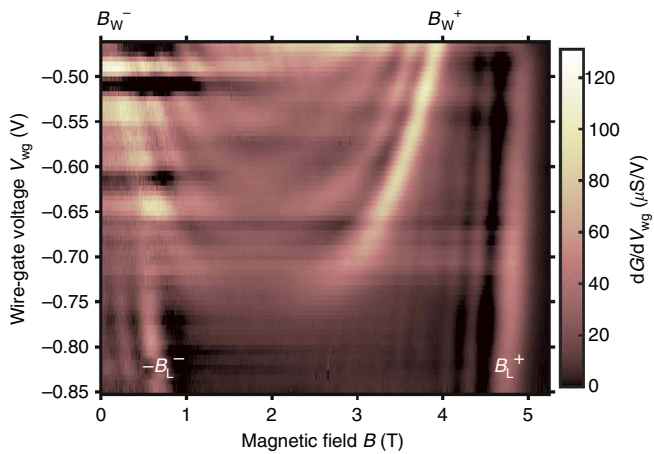

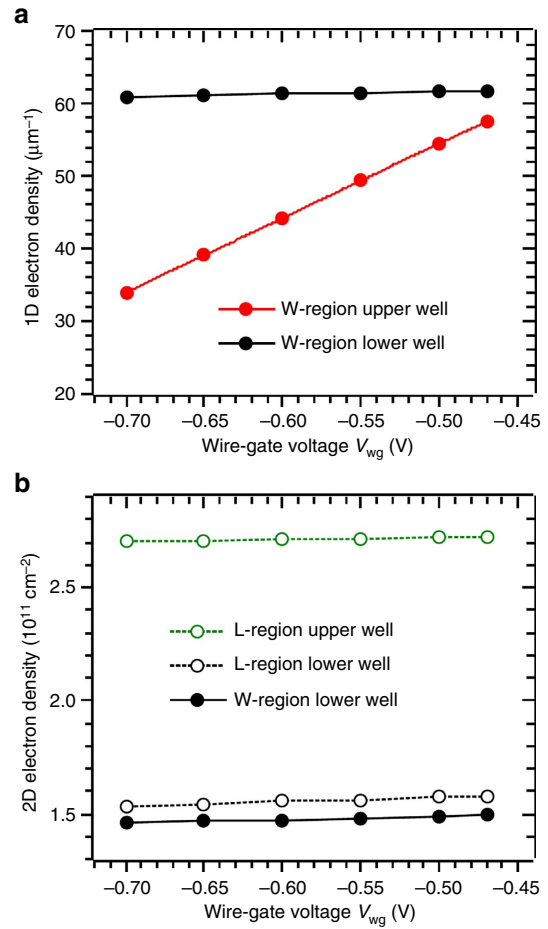

**Figure 3 | 1D wire subbands participating in 1D–2D tunnelling.** The derivative of the tunnelling conductance $G$, with respect to the wire–gate voltage $V_{wg}$, is plotted as a function of $V_{wg}$ and in-plane magnetic field $B$ perpendicular to the wires, for conditions close to equilibrium ($V_{dc} \approx 0$). Each quasi-parabolic resonance corresponds to one of the 1D wire subbands. The centres of the tunnelling resonances approximately correspond to the inflection points in the colour contrast, in going upwards from bright to dark. The magnetic field values $B_W^-$, $B_W^+$, $-B_L^-$ and $B_L^+$, at which tunnelling resonances cross the $V_{dc} = 0$ axis in $G(B, V_{dc})$ maps, are indicated. $B_W^-$ and $B_W^+$ correspond to tunnelling in W (wire) regions of the device. $-B_L^-$ and $B_L^+$ correspond to tunnelling in L (lead) regions.

Note that the particle and hole sectors for the upper well are inverted relative to those for the lower well. The width of 1D bands in Fig. 5 is to a large extent associated to intrinsic physical effects. The width parallel to the magnetic-field axis correlates to momentum uncertainty, which (according to the Heisenberg's principle) is inversely proportional to the wire length. Wire shortness is thus a cause of resonance broadening. 1D bands are also broad parallelly to the dc-bias axis, because of the power-law (slow-decaying) shape of 1D singularities. Tunnelling maps are symmetric with respect to the sign of the magnetic field. Note, however, that the map in Fig. 5 is not symmetric with respect to the sign of the dc-bias. The dispersions of the s− and s+ bands are symmetric, but this is just a local symmetry. The map does not exhibit general symmetry with respect to $V_{dc}$. This indicates that potentials are different in the upper and lower layers. Note that resonances in the map of Fig. 5 do not cross the $V_{dc} = 0$ axis at $B = 0$. That is, $B_{W,L}^- \neq 0$. This indicates that equilibrium electron densities, and Fermi wavenumbers, are different in the upper and lower layers.

**Regime of three 1D wire subbands filled.** Maps corresponding to the regime of three 1D subbands filled ($V_{wg} = -0.55$ V) are shown in Fig. 6. To better discriminate the location of resonances, the derivatives $dG/dV_{dc}$ and $dG/dB$ are plotted. Mapping of $dG/dB$ ($dG/dV_{dc}$) emphasizes conductance modulations as a function of momentum (excitation energy). In Fig. 6, tunnelling resonances correspond to peaks of enhanced conductance, across which colour contrast changes quite abruptly. These peaks define quasi-parabolic lines on the maps (see the guide-to-the-eye lines drawn). Note that valleys, with colour contrast sequence opposite to that of peaks, do not correspond to resonances, but to regions of low conductance. The resonances marked by the green lines in Fig. 6a,b have the same meanings as in Fig. 5. Most importantly here, the black lines mark 1D UW-wire bands. The dashed-black line drawn in Fig. 6a,b for $V_{dc} > 0$ (lower half of the plots) marks the (first) spinon hole subband (s−), best discriminated in

**Figure 4 | Equilibrium electron densities in the upper and lower wells.** (**a**) 1D and (**b**) 2D electron densities in the upper (red and green lines and symbols) and lower (black lines and symbols) wells, for the wire (W, solid lines and symbols) and lead (L, dashed lines and open symbols) regions of the device, as a function of the wire–gate voltage $V_{wg}$, estimated from the measured magnetic field values $B_W^-$, $B_W^+$, $-B_L^-$ and $B_L^+$. Note that electrons are not laterally confined in the lower well, and so the W-region lower-well density that has physical meaning is the 2D-like one, shown in **b**. A 1D-like W-region lower-well density is shown in **a** just for comparison with the 1D-like density corresponding to the upper-well wires.

Fig. 6a,c as a line with red/blue contrast. Remarkably, a band with symmetric (inverted) dispersion is seen in the particle sector (dashed-black line for $V_{dc} < 0$, upper half of the plots), as in the single-subband regime (s+). The inverted band is best discriminated in Fig. 6b,d as a line with red/white contrast. The (first) holon subband (h+, marked by the solid-black line) appears well discriminated in the particle sector ($V_{dc} < 0$) at high energies, but hardly discernible in the hole sector ($V_{dc} > 0$). In addition, Fig. 6b,d reveals quasi-horizontal features. Similar states are seen in the regime of fully depleted wires. The quasi-horizontal features seen in Fig. 6b,d possibly are superpositions of contributions from (i) the bottoms of the second and third UW wire subbands, and from (ii) localized states[29–31] likely formed at intersection and/or bending points of the UW leads (Lb, Lc and/or Ld regions in Fig. 1c). Momentum-conserving resonances in Figs 5 and 6 (green and black lines) are seen not to be perfectly symmetric in their right and left sides, as opposed to the symmetric dispersions to which they correspond. The small asymmetry is known to be caused by capacitance effects[32], for which we account on the basis of a simple model previously used[15] (see Methods section for details).

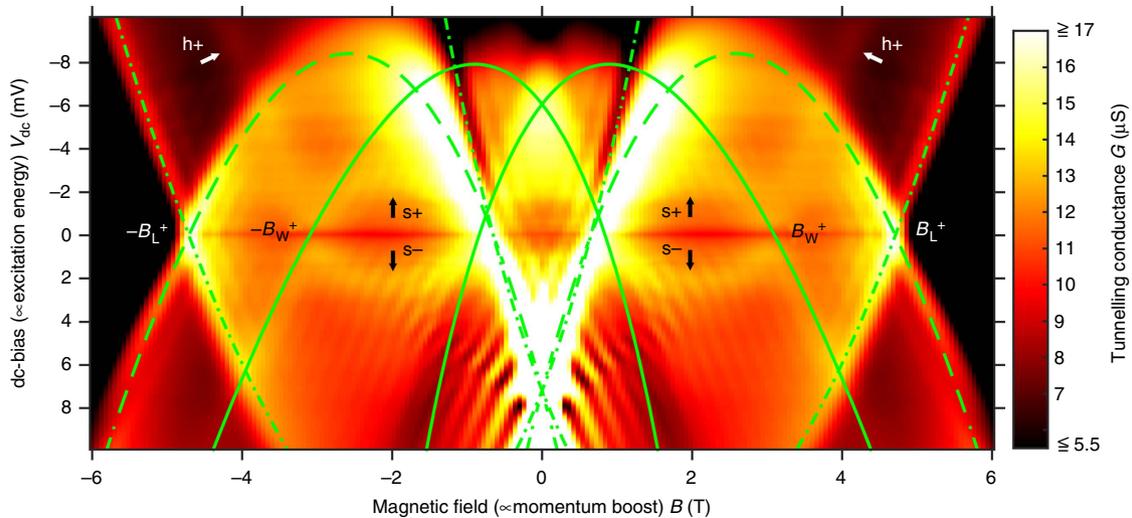

**Figure 5 | Regime of a single 1D wire subband filled.** The tunnelling conductance $G$ is plotted as a function of the dc-bias $V_{dc}$ and in-plane magnetic field $B$ perpendicular to the wires, for the regime of a single 1D subband filled, achieved by setting the wire–gate voltage to $V_{wg} = -0.69$ V. The electron density in the wires is $n_{1D} \cong 35\ \mu m^{-1}$. The solid (dashed) green lines mark resonances corresponding to tunnelling between upper-well wire W (lead L) ground states and lower-well states; they reveal the dispersion of the elementary excitations in the 2D lower well, in the wire (lead) regions of the device. The dash-dotted green lines mark resonances corresponding to tunnelling between LW ground states and UW lead states; they reveal the dispersion of the elementary excitations in the 2D UW leads. The resonances marked by the black and white arrows correspond to tunnelling from LW ground states to UW wire states; they reveal dispersions of elementary excitations in the 1D UW wires: a holon band in the particle sector (h+), and spinon bands in the hole (s−) and particle (s+) sectors. The labels $\pm B^+_{W,L}$ indicate specific magnetic field values at which W and L tunnelling resonances cross the $V_{dc} = 0$ axis.

**Conductance oscillations.** When tunnelling occurs through a barrier of finite (short) length $l$ in the in-plane direction perpendicular to the magnetic field, momentum is not strictly conserved in tunnelling and, as a consequence, the tunnelling conductance exhibits oscillations[32–35]. The maps in Figs 5 and 6 exhibit oscillations of this kind (diagonal short-period conductance oscillations), associated to tunnelling in (central or peripheral) La regions of the device (see Fig. 1c). In these regions, the UW states are laterally confined under the action of the wire gates, in the in-plane direction perpendicular to the magnetic field. For square confinement, the magnetic-field period $\Delta B$ of tunnelling conductance oscillations has been shown[34,35] to be $\Delta B \approx \phi_0/(ld)$, where $\phi_0 = 2\pi\hbar/e$ is the quantum of flux. For soft confinement, the period of the oscillations has been shown[34,35] to be $\Delta B \approx \phi_0/(\Delta x\, d)$, where $\Delta x$ is the distance between the classical turning points. In our device, La regions are nominally 0.6 μm-wide, corresponding to the gap between the wire gates. In the maps of Figs 5 and 6, the oscillation period is $\Delta B \approx 0.26$ T, which corresponds to a confinement width $\Delta x \approx 0.5$ μm. This is just a bit smaller than the gap between the wire gates, consistent with the small lateral depleting effect that the gates produce.

**Band assignment.** Within the linear theory of Luttinger liquids, the spectral function of repulsive spinful 1D fermions has power-law singularities at linear spinon and holon mass shells $\varepsilon^{LL}_{s,c}(k) = v^F_{s,c}(\pm k - k^{1D}_F)$ and at the inverted holon mass shell $-\varepsilon^{LL}_c(k) = -v^F_c(\pm k - k^{1D}_F)$ near the right (+) and left (−) Fermi points[22,23]. Spinons and holons travel at different velocities $v^F_s$ and $v^F_c$, respectively ($v^F_s < v^F_c$). Within the nonlinear theory[22–24], the excitation with lowest possible energy for a given momentum (edge of support) is predicted to coincide, in the main $|k| < k^{1D}_F$ region of the hole sector ($\omega < 0$), with the spinon mass-shell $\varepsilon_s(k)$. In the main region of the particle sector ($\omega > 0$), the edge is predicted to coincide with the inverted spinon mass shell $-\varepsilon_s(k)$. For arbitrary $k$ in the regions $(2j-1)k^{1D}_F < k < (2j+1)k^{1D}_F$ defined by integer $j$, the edge is predicted to coincide with periodic $2jk^{1D}_F$-shifts of the functions $\pm\varepsilon_s(k)$, see Fig. 1a in ref. 22. Holon singularities may exist at the holon mass shell $\varepsilon_c(k)$, as well as at $2jk^{1D}_F$-shifted holon lines. The leading nonlinear correction to the holon dispersion near the Fermi points is quadratic[23,24]:

$$\varepsilon_c(k) \approx v^F_c\hbar(\pm k - k^{1D}_F) + \frac{\hbar^2}{2m^*_c}(\pm k - k^{1D}_F)^2$$

$$= \frac{\hbar^2}{2m^*_c}\left[k^2 - (k^{1D}_F)^2\right], \quad (6)$$

whereas that to the spinon dispersion is cubic:

$$\varepsilon_s(k) \approx v^F_s\hbar(\pm k - k^{1D}_F) - \xi(\pm k - k^{1D}_F)^3. \quad (7)$$

Here, $m^*_c = \hbar k^{1D}_F/v^F_c$ is the holon effective mass, and $\xi$ is a positive parameter. Note that the nonlinear corrections have opposite sign for holons and for spinons in the particle sector ($|k| > k^{1D}_F$), and same sign in the hole sector ($|k| < k^{1D}_F$). The precise shapes of the spinon and holon dispersions, away from the Fermi points, are generally unknown theoretically.

According to the nonlinear theory[22,23], states at the edge of support are spinon states (not holon states); spinon excitations are protected from decay, but holon excitations are subjected to decay. Hence, the resonance labelled s− in Fig. 5 (as well as that marked by the dashed-black line in the lower half of Fig. 6a,b) should correspond to the main spinon band. Consistent with the nonlinear theory, the symmetric inverted replica seen in the particle sector (s+ in Fig. 5, and the upper-half dashed-black-line resonance in Fig. 6a,b) appears to be its shadow (spinon) band. On the other hand, given its positive curvature (upwards bending), the resonance labelled h+ in Fig. 5 (as well as that marked by the solid-black line in the upper half of Fig. 6a,b) should correspond to a holon band. It cannot correspond to a spinon band, because the dispersion of spinons has opposite curvature in this sector (see Fig. 1a in ref. 22).

**Spinon and holon effective masses.** To extract quantitative information, we simulate the experimental holon dispersion

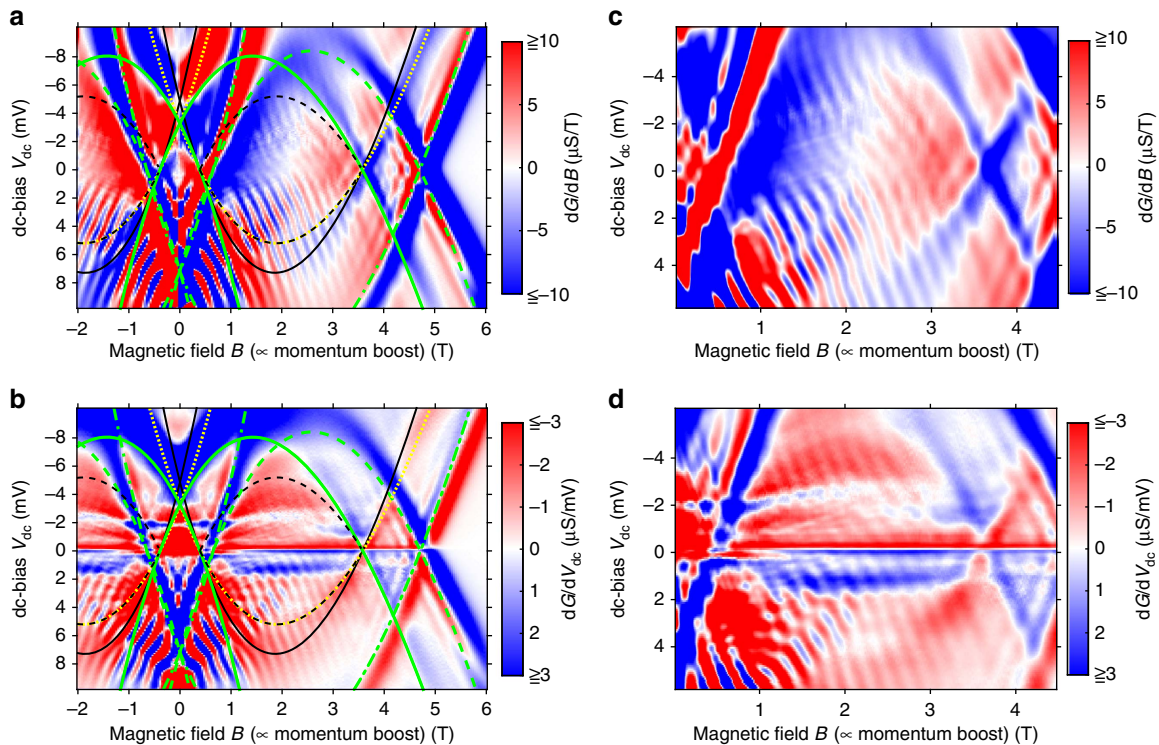

**Figure 6 | Regime of three 1D wire subbands filled.** The derivatives of the tunnelling conductance $G$, with respect to either $B$ or $V_{dc}$, are plotted as a function of the dc-bias $V_{dc}$ and in-plane magnetic field $B$ perpendicular to the wires, for the regime of three 1D wire subbands filled, achieved by setting the wire–gate voltage to $V_{wg} = -0.55$ V. The electron density in the wires is $n_{1D} \cong 49\,\mu m^{-1}$. The resonances marked by the solid and dashed green lines reveal the dispersion of the elementary excitations in the 2D lower well, in the W (wire) and L (lead) regions of the device, respectively. The resonances marked by the dash-dotted green lines reveal the dispersion in the 2D upper-well leads. The resonances marked by the black lines reveal (first subband) dispersions for elementary excitations in the 1D upper-well wires, for spinons (dashed lines) and for holons (solid lines). (**a**,**b**) Wide $dG/dB$ and $dG/dV_{dc}$ maps. (**c**,**d**) High-resolution $dG/dB$ and $dG/dV_{dc}$ maps, zooming in on the magnetic-field region $B_W^- \leq B \leq B_W^+$, corresponding to the main wavenumber region $|k| \leq k_F^{1D}$ for the 1D upper-well wires.

(h+, solid-black lines) with a quadratic functional form (equation (6)). It has been shown[26] that the spinon dispersion in the nonlinear regime is approximately parabolic. Therefore, we simulate the spinon dispersion (s−, dashed-black lines) in the hole sector (lower half of the plots) of the main $|k| < k_F^{1D}$ region ($B_W^- < B < B_W^+$) with a quadratic functional form, analogous to that of holons, but with a different effective mass $m_s^* = \hbar k_F^{1D}/v_s^F$:

$$\varepsilon_s(k) \approx v_s^F \hbar\big(\pm k - k_F^{1D}\big) + \frac{\hbar^2}{2m_s^*}\big(\pm k - k_F^{1D}\big)^2$$

$$= \frac{\hbar^2}{2m_s^*}\big[k^2 - (k_F^{1D})^2\big]. \quad (8)$$

We express the holon and spinon effective masses as $m_{c,s}^* = \tilde{K}_{c,s} m_{2D}^*$, where $\tilde{K}_{c,s}$ are phenomenological parameters, accounting for renormalization of the effective mass due to 1D confinement[26], and $m_{2D}^*$ is the effective mass of the noninteracting electron-like quasiparticles in the parent UW Fermi liquid. Note that a change in the nature of the elementary excitations occurs upon imposing lateral 1D confinement, but not upon 2D confinement. $\tilde{K}_{c,s}$ thus account for mass renormalization in going from a 2D Fermi liquid to a 1D Luttinger liquid. The spinon dispersion is found to be characterized by an effective mass similar to that of noninteracting quasiparticles in the UW-lead and LW Fermi liquids, that is, $m_s^* = m_{2D}^*$, corresponding to $\tilde{K}_s = 1$. (See Methods section for details of simulations of the LW and UW-lead dispersions.) Note that the spinon momentum is bounded[23], and so the spinon quasi-parabolas (lower-half dashed-black lines in Fig. 6) do not continue beyond $B_W^- < B < B_W^+$ ($|k| < k_F^{1D}$), but theoretically replicate through shifts and inversions (see Fig. 1a in ref. 22). In contrast, the holon momentum is not bounded[23], and so the holon quasi-parabolas (solid-black lines in Fig. 6) theoretically extend from the particle sector $\omega > 0$ ($V_{dc} < 0$) into the hole sector $\omega < 0$ ($V_{dc} > 0$) without changing the sign of their curvature (see Fig. 1a in ref. 22). Fitting of the holon dispersion is carried out in the particle sector (upper half of the plots), where it is well discriminated. The holon dispersion is found to be characterized by an effective mass notably smaller than that of noninteracting quasiparticles in the UW-lead and LW Fermi liquids. The mass renormalization parameter is found to be $\tilde{K}_c = 0.65$ ($\tilde{K}_c = 0.75$) in the regime of one (three) subband(s) filled. This is consistent with the expected effect of repulsive interactions, for which the theoretical Luttinger-liquid parameter $K_c$ is in the range $0 < K_c < 1$. The higher value of $\tilde{K}_c$ in the regime of three subbands filled is understood considering that the second and third subbands somewhat screen interactions in the first subband. Nevertheless, observation of the inverted spinon shadow band (s+), when more than one subband is filled (Fig. 6), indicates that the first 1D subband remains highly correlated even in the presence of screening by higher subbands. We emphasize that the wire bands seen in the hole sector (s− in Fig. 5, lower-half dashed-black lines in Fig. 6a,b) and in the particle sector (h+ in Fig. 5, upper-half solid-black lines in Fig. 6a,b) cannot be adequately described with a single parabola. A continuation of the lower-half dashed-black lines—dotted yellow lines in Fig. 6a,b—is clearly unsatisfactory in fitting the resonance observed in the particle sector at high energies (upper half of the plots, $V_{dc} \ll 0$). The spinon (s−, lower-half dashed-black-line) and holon (h+, upper-half solid-black-line) bands are characterized by notably different effective masses.

## Discussion

We finally comment on the physical meaning of the observed 1D bands in the context of the new nonlinear theory[22–24]. The injection of a spin-up (spin-down) electron into a wire leads to the creation (annihilation) of a spinon and the creation of a holon. If the energy of the incoming electron is $\hbar\omega = \varepsilon_s(k)$, the final state will contain a spinon carrying the whole-energy $\hbar\omega$, and a holon carrying no energy. Similarly, if the energy of the incoming electron is $\hbar\omega = \varepsilon_c(k)$, the final state will contain a holon with energy $\hbar\omega$, and a spinon with no energy. Spinon excitations are protected from decay by energy and momentum conservation laws. Therefore, the (spinon) edge singularity is predicted to be robust. Holon excitations are subjected to decay to some (not accurately known) extent, and so it is theoretically unclear whether holon singularities should be observable away from the Fermi points[23,24]. Here, the holon band appears experimentally well discriminated in the particle sector, but hardly discernible in the hole sector. That is, holon particles appear long-lived, but holon holes short-lived. Phenomenological theoretical expressions yield the $j$- and $k$-dependent exponents of the power-law singularities at the spinon and holon mass shells[22–24]. The spinon exponents in the particle sector are similar to those in the hole sector, except for different selection rules $m_{\pm}$. Hence, the (main region) spinon singularity is theoretically predicted to be a divergent one in the hole sector (thus easy to observe), and a convergent one in the particle sector (thus difficult to observe). This would explain why observation of the spinon shadow band in the particle sector of the main region, reported here, has been previously elusive. Our results indicate that this shadow band does exist, supporting the predictions of the new nonlinear theories[22–24]. They show that the shadow band has appreciable weight for short wires ($\sim 1\,\mu$m long), wire length appearing as a relevant parameter modulating the spectral weight of shadow bands for electronic (spinful) systems. Theoretical studies[27,28] indicate that the spectral weight of shadow bands for spinless systems increases strongly with decreasing wire length. Here, the experimental observation of the inverted spinon shadow band (s+) appears to be facilitated by (i) the wire shortness, (ii) the array design of our device that boosts the wire signal and (iii) the homogeneity of the lithography.

## Methods

**Experimental device.** The vertical structure of the device, grown by molecular-beam epitaxy, comprises two 18 nm-wide GaAs quantum wells, a 14 nm-wide tunnel barrier in between, and spacer and Si-doped $Al_{0.33}Ga_{0.67}As$ layers at both sides. The structure terminates with a 10 nm-wide GaAs cap layer. The tunnel barrier is a $10\times[(0.833\,\text{nm})Al_{0.33}Ga_{0.67}As/(0.556\,\text{nm})GaAs]$ superlattice. The lower (upper) spacer is 40 nm (20 nm) wide and the Si-doped layers ($10^{18}$ donors cm$^{-3}$) are 40 nm wide. The electrical (surface) structure of the device was fabricated on a 200 μm-wide Hall bar (Fig. 1b). Electron-beam lithography was used to pattern surface gates, that is, split (SG), mid-line (MG), bar (BG) and cut-off (CG) gates (Fig. 1a)—used to define the experimental area and to set-up tunnelling conditions—and a $30\times 200$ array of air-bridged wire gates (Fig. 1d and Fig. 2)—used to define the quantum wires in the upper well. The wire gates are 1 μm-long and 0.3 μm-wide. They are separated by 0.19 μm-wide gaps (W wire regions) in the transverse direction and by 0.6 μm-wide gaps (L lead regions) in the longitudinal direction (Fig. 1c). The device dimensions were chosen carefully to achieve minimal modulation of the LW carrier density by the negative wire–gate voltage.

**Tunnelling set-up.** The two quantum wells are separately contacted with AuGeNi Ohmic contacts by using a surface-gate depletion technique. At one side of the device (Fig. 1a), gate SG is negatively biased to pinch off both the upper and lower layers, while gate MG is positively biased to open a narrow conducting channel in the upper layer only. At the other side of the device, gates BG and CG are negatively biased to pinch off just the upper layer. Hence, the current injected through one of the Ohmic contacts is forced to tunnel between the upper and lower layers, before flowing out through the other Ohmic contact.

**Modelling of the LW and UW-lead dispersions.** The electron density in the lower well (UW leads) is $n_{LW} = 1.5\times 10^{11}$ cm$^{-2}$ ($n_{UW,L} = 2.7\times 10^{11}$ cm$^{-2}$), obtained from the crossing magnetic-field values $B_L^{\pm}$. For these densities, 2D electron systems are known to be Fermi liquids with effective mass renormalized by interactions[36,37]. We thus simulate the LW and UW-lead dispersions with the parabolic functional form

$$\varepsilon_{2D}(k) = \frac{\hbar^2}{2m_{2D}^*}\left[k^2 - (k_F^{2D})^2\right], \quad (9)$$

characteristic of Fermi-liquid quasiparticles, where $m_{2D}^*$ is the effective mass in the lower well or in the UW leads. Replacing the corresponding dispersions in equation (5), one obtains functions of the type

$$V_{dc}(B) = \pm\frac{ed^2}{2m_{2D}^*}B^2 + c_1 B + c_0 \quad (10)$$

that we use to simulate the tunnelling resonances corresponding to the LW (+) and UW-lead (−) dispersions. By setting $d = 32$ nm (nominal value), the LW and UW-lead dispersions are found to be well described by an effective mass $m_{2D}^* = 0.050 m_0 = 0.75 m_b$, somewhat smaller than the band mass of GaAs ($m_b = 0.067 m_0$), where $m_0$ is the electron rest mass. This is consistent with previous measurements and theoretical predictions for quantum wells with similar densities[38–42]. If the bands were substantially bent across the quantum wells, the wavefunctions would be somewhat displaced relative to flat-band conditions, and the distance $d$ (between the centres of the UW and LW wavefunctions) would somewhat differ from the distance between the centres of the quantum wells (32 nm). Equally good fits are obtained if we impose the effective mass to be equal to the band mass ($m_{2D}^* = m_b$) and allow for variation of $d$. The best fit is then achieved for $d = 37$ nm. We cannot determine the individual contributions of $m_{2D}^*$ and $d$ to the curvature of tunnelling resonances. Nevertheless, we note that the mass $m_{2D}^* = 0.050 m_0$ is in very good agreement with the careful measurements carried out by Hayne et al.[38,39].

**Modelling of capacitance effects.** Tunnel transport between the upper and lower layers of the device, across its dielectric barrier, is affected by capacitance effects, which cause a small deformation of the observed dispersions. The finite capacitance of the device ($C/A$ per unit area, $C/L$ per unit length) leads to a small increase/reduction of the electron density $\pm\delta n_{2D/1D}$ at each side of the barrier ($e\delta n_{2D} = V_{dc}C/A$ for 2D systems, and $e\delta n_{1D} = V_{dc}C/L$ for 1D systems), and consequently to changes of the Fermi wavenumbers of the source and drain systems ($\pm\delta k_F^{2D} = \pm\delta n_{2D}/k_F^{2D}$ for 2D systems, and $\pm\delta k_F^{1D} = \pm\delta n_{1D}/2$ for 1D systems). To account for capacitance effects[15] in simulations of the measured resonances, we replace the Fermi wavenumbers, corresponding to zero capacitance, by modified ones

$$k_F^{2D\prime} = k_F^{2D} \pm \delta k_F^{2D} = k_F^{2D} \pm \frac{\pi\,V_{dc}\,C}{e\,k_F^{2D}\,A} \quad (11)$$

for 2D systems, and

$$k_F^{1D\prime} = k_F^{1D} \pm \delta k_F^{1D} = k_F^{1D} \pm \frac{\pi\,V_{dc}\,C}{2e\,L} \quad (12)$$

for 1D systems, in the expressions describing tunnelling conditions (equation (5)). Hence, dispersions $V_{dc}(B)$ initially symmetric become somewhat asymmetric. Satisfactory fits are achieved with capacitance values $C/A \approx 0.004$ F m$^{-2}$, $C/L \approx 0.1$ nF m$^{-1}$.

**Data availability.** Data associated with this work are available at the University of Cambridge data repository (http://dx.doi.org/10.17863/CAM.797; see ref. 43).

### Acknowledgements
This work was supported by the UK EPSRC [Grant Nos. EP/J01690X/1 and EP/J016888/1]. M.M. thanks J. Waldie for her assistance with device fabrication and transport measurements, and P. See for his advice on device fabrication.

### Author contributions
Project planning: C.J.B.F. and A.J.S.; MBE growth: I.F. and D.A.R.; optical and electron-beam lithography: M.M. and J.P.G., with assistance from Y.J. and G.A.C.J.; transport measurements: M.M., C.J.B.F. and Y.J.; analysis of results and theoretical interpretation: M.M., C.J.B.F., O.T. and A.J.S.; manuscript preparation: M.M. with contributions from C.J.B.F.

### Additional information

**How to cite this article:** Moreno, M. et al. Nonlinear spectra of spinons and holons in short GaAs quantum wires. *Nat. Commun.* 7:12784 doi: 10.1038/ncomms12784 (2016).

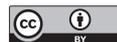